\documentstyle[aps,prl,twocolumn,floats]{revtex}
\begin{document}
\draft
\title{{\rm PHYSICAL PAPER}\hfill {\sl Version of \today}\\~~\\
Pole solutions in the case of problems of flame front propagation
and Saffman-Teylor "finger" formation without surface tension:
open problems and possible ways of their solutions.}
\author { Oleg Kupervasser }
\address{ Hi-tech\\
Oplus Ltd. Israel}\maketitle
\narrowtext
\begin{abstract}
Some physical problems as flame front propagation or Laplacian
growth without surface tension have nice analytical solutions
which replace  its complex integro-differential motion equations
by simple differential equations of poles motion in a complex
plane. Investigation of these equation was the main topic of
Kupervasser Oleg Ph.D. Thesis[7]. Some very interesting open
problems were hanged up there.Here we give these open problems and
possible ways of their solutions.
\end{abstract}
%\leftskip 54.8pt
%\rightskip 54.8pt
%}]
%\widetext
%%%%%%%%%%%%%%%%%%%%%%%%%%%%%%
In the beginning let us to consider the case of Saffman-Teylor
"finger" formation.

1)   The case of Laplacian growth in the channel without surface
tension was in details considered by Mark Mineev-Weinstein and
Dawson [1]. In this case the problem has the beautiful analytical
solution. Moreover they assumed that all major effects in the case
with vanishingly small surface tension may be received also
without surface tension. It would allow applying to vanishingly
small surface tension case the powerful analytical methods
developed for the no surface tension case. However without
additional assumptions this hypothesis may not be accepted.

The first objection is related to finite time singularities for
some initial conditions. Actually, for overcoming this difficulty
the regular item with surface tension was introduced. This surface
tension item is resulting in loss of the analytical decision.
However regularization may be carried out much more simply -
simply by rejecting the initial conditions which result to these
singularities.

The second objection is given in work Siegel and Tanveer [2].
There it is shown, that in numerical simulations in a case with
any (even vanishingly small) surface tension any initial thickness
"finger" extends up to $1 \over 2$ thickness width of the channel.
The analytical solution in a case without a surface tension
results in constant thickness of the "finger" equal to its initial
size that may be arbitrary. Siegel and Tanveer however did not
take into account the simple fact, that numerical noise introduces
small perturbation or to the initial condition, or even during
"finger" growth, which is equivalent to the remote poles, and with
respect to this perturbation the analytical solution with constant
"finger" thickness is unstable. By Mark Mineev-Weinstein [3] it
was shown, that similar pole perturbations can give, at the some
initial conditions, extending up to the Siegel and Tanveer
solutions. This positive aspect of the paper [3] was mentioned by
Sarkissian and Levine in them comment [4]. Summing up, it is
possible to tell, that for identity of the results with and
without surface tension it is necessary to introduce a permanent
source of the new remote poles:  it may be either external noise
or infinite number of poles in an initial condition. What from
these methods is preferred it is a open question yet. In the case
of flame front propagation it was shown [7], that external noise
is necessary for an explanation of flame front velocity increase
with the sizes of system: the infinite number of poles in an
initial condition can not give this result. It is interesting to
know, what is situation in the channel Laplacian growth.

One of main results of Laplacian growth in the channel with a
small surface tension is Saffman-Teylor  "finger" formation with
the thickness equal to $1 \over 2$ thickness of the channel. And
to use the analytical result received for zero surface tension, it
is necessary to prove, that formation of the "finger" with
thickness equal to $1 \over 2$ thickness of the channel takes
place without surface tension also. In our teamwork with Mark
Mineev-Weinstein [5] it was shown, that for finite number of poles
at almost all allowed (in the sense of not approaching to finite
time singularities) initial conditions, except for small number of
some degenerated initial conditions, they  have asymptotic as some
"finger" with any possible thickness. It should be mentioned, that
the solutions and asymptotic found in [5] for finite number of
poles are though also idealization, but quite have real sense for
any finite intervals of time between appearance of   the new poles
introduced into system by external noise or connected to an
entrance to the system of remote poles of an initial condition,
including infinite number of such poles. The theorem proved in [5]
and may be again applied for this final set of new and old poles
is again received asymptotic, being again "finger", but already
with possible new, distinct from former, thickness. Thus,
introduction of a source of new poles results only in possible
drift of thickness of the final "finger", but not changing of type
of this solution. It should be mentioned, that instead of
periodical boundary conditions, much more realistic physical
boundary conditions may be introduced [6], forbidding a stream
through a wall which insert additional, probably useful,
restrictions on a positions, number and parameters of new and old
poles (explaining, for example, why the sum of all complex
parameters $\alpha_i$ for poles give the real value $\alpha$ for
the pole solution (5) in [3]), not influencing, however, as shown
in [7], on correctness and applicability proved in [5] results and
methods of their including. Given in [3] by Mark Mineev-Weinstein
"proof", that steady asymptotic for Laplacian growth in a channel
with zero surface tension is single "finger" with thickness equal
to $1 \over 2$ thickness of the channel, is unequivocally
erroneous: completely the same method which was used in [3] to
prove and demonstrate instability of  "finger" with thickness
distinct from $1 \over 2$ with respect to introducing the new
remote poles, instability of "finger" with thickness equal to $1
\over 2$ may be proved and demonstrated! This objection was
repeatedly stated to Mark Mark Mineev-Weinstein before the
publication of his paper [3], however has not found any answer
there. Moreover, in our teamwork [5] was is shown, that for finite
number of poles any thickness "finger" is possible as asymptotic.
It does not mean, nevertheless, that privileged role of  "finger"
with thickness $1 \over 2$ cannot be proved in the case of surface
tension absence, but means only that such the proof are not given
in [3]. Let us try to give these correct arguments here. The
general pole solution (5) in work [3] is characterized by the real
parameter $\alpha$ being the sum of the complex parameters
$\alpha_i$ for poles. Thickness of the asymptotic finger is simple
function of $\alpha$: ($Thickness =1- {\alpha \over 2}$). The
value ($\alpha =1$) corresponds to thickness $1 \over 2$. As far
as possible thickness of the "finger" is between 0 and 1, possible
$\alpha$ value is in an interval between 0 and 2: ($0 < \alpha <
2$). The value $\alpha =1$ corresponding to thickness $1 \over 2$
is exactly in the middle of this interval. What occurs to quite
possible initial pole conditions with $\alpha$ outside of limits
from 0 up to 2? They are "not allowed" because of already known to
us finite time singularities [5]. Also a part of solutions inside
of interval $0 < \alpha < 2$ results to the similar finite time
singularities. Exact necessary conditions, whether defining the
initial pole condition as "allowed", i.e. singular, is still a
open problem.
 How number of these "allowed" initial pole conditions (to be exact speaking,
 their percent from the full number of the possible initial pole conditions
 corresponding to the given real value $\alpha$)  is distributed inside of this interval?
 From the reasons of a continuity and symmetry with respect to $\alpha =1$ it is possible to
 conclude, that this distribution has a minimum in point $\alpha =1$  (thickness $1 \over 2$!), the
 value which is the most remote from both borders of interval $0 < \alpha < 2$, being increased
 to borders $\alpha =2$ or 0, and reaching 100 %
 from all pole solutions outside of these borders. I.e. thickness $1 \over 2$
 is the most probable because for this thickness value the minimal percent of
 potentially capable to give such thickness value initial conditions is "not allowed",
 i.e. results to singularities.
Source of new poles results to the drift of finger thickness, but
this thickness drift is closed to the most probable and average
size equal to $1 \over 2$! The similar result is obtained in the
case of Saffman-Teylor "finger" with vanishingly small surface
tension and with some external noise. As it was desirable to be
proved. Similar idea, that the initial conditions resulting to
singularities, can provide the selected and special role of
thickness $1 \over 2$ solution stated by Procaccia [8] also. These
given arguments are not, certainly, the strict proof, but only
specifies a way to it. The inquisitive reader is invited to make
and publish it.

Let us pass to a problem of flame front propagation.

         2) For a cylindrical case of the flame front propagation problem at absence of noise (only numerical noise) (look [7] and the bibliography there) by Sivashinsky with help of numerical methods it was shown, that the flame front is continuously accelerated. During all this account time it is not visible any attributes of saturation. To increase time of the account is a difficult task. Hence, absence or presence of velocity saturation in a cylindrical case, as consequence of the flame front motion equation it is still a open problem yet.

For the best understanding of dependence of flame front velocity
as functions of its radius in a cylindrical case similar
dependence of flame front velocity on width of the channel (in a
flat case) also was analyzed by numerical methods. Growth of
velocity is also observed and at absence of noise (only numerical
noise!) also any saturation of the velocity it is not observed.
Introduction obvious Gaussian noise results to appearance of a
point of saturation and its removal from the origin of coordinates
with decreasing of noise amplitude, allowing extrapolating results
on small numerical noise. ( Fig. 2.6 in [7])

Hence, introducing of Gaussian noise in numerical calculation also
for a cylindrical case can again results to appearance of a
saturation point and will allow to investigate its behavior as
function of noise amplitude by extrapolating results on small
numerical noise. The Inquisitive Reader loved by me is invited
again to make it and to publish the received interesting results.

\end{document}